\documentclass[draft,jgrga]{agu2001}

\ifgalley
\let\eject\relax
\fi

\usepackage{graphicx}
\usepackage{latexsym}

\authorrunninghead{MUTEL ET AL.}
\titlerunninghead{Striated AKR: A REMOTE TRACER OF ION HOLES}

\authoraddr{R. L. Mutel, Dept. Physics and Astronomy, University of Iowa,
Iowa City IA 52242 (robert-mutel@uiowa.edu)}
\authoraddr{J. D. Menietti, Dept. Physics and Astronomy, University of Iowa,
Iowa City IA 52242 (john-menietti@uiowa.edu)}
\authoraddr{I.W. Christopher, Dept. Physics and Astronomy, University of Iowa,
Iowa City IA 52242 (ivar-christopher@uiowa.edu)}
\authoraddr{D. A. Gurnett, Dept. Physics and Astronomy, University of Iowa,
Iowa City IA 52242 (donald-gurnett@uiowa.edu)}
\authoraddr{J.M. Cook, Dept. Physics and Astronomy, University of Iowa,
Iowa City IA 52242 (justin-cook-2@uiowa.edu)}

\begin{document}
\def\etal{et al.\ }

\title{Striated AKR Emission: A Remote Tracer of Ion Solitary Structures}

\authors{
R.L. Mutel, \altaffilmark{1}
J .D. Menietti,  \altaffilmark{1}
I. W. Christopher,  \altaffilmark{1}
D. A. Gurnett,  \altaffilmark{1}
J. M. Cook  \altaffilmark{1}
}
\altaffiltext{1}
{Dept. Physics and Astronomy, University of Iowa, Iowa City IA 52242, USA.}

\begin{abstract}

We describe the statistical properties of narrowband drifting auroral
kilometric radiation ('striated' AKR) based on observations from the
Cluster wideband receiver during 2002-2005. We show that the observed
characteristics, including frequency drift rate and direction, narrow
bandwidth, observed intensity, and beaming angular sizes are all
consistent with triggering by upward traveling ion solitary structures
(`ion holes'). We calculate the expected perturbation of a
horseshoe electron distribution function by an ion hole by integrating the
resonance condition for a cyclotron maser instability (CMI) using the
perturbed velocity distribution. We find that the CMI growth rate can be
strongly enhanced as the horseshoe velocity distribution contracts inside
the passing ion hole, resulting in a power gain increase greater than 100
dB.  The gain curve is sharply peaked just above the R-mode cut-off
frequency, with an effective bandwidth $\leq$50 Hz, consistent with the
observed bandwidth of striated AKR emission. Ion holes are observed
\textit{in situ} in the acceleration region moving upward
with spatial scales and speeds consistent with the observed bandwidth and
slopes of SAKR bursts. Hence, we suggest that SAKR bursts are a remote
sensor of ion holes and can be used to determine the frequency of
occurrence, locations in the acceleration region, and lifetimes of these
structures.  \end{abstract}

\begin{article}
\section{Introduction}
Auroral kilometric radiation (AKR) bursts exhibit a wide variety of fine
structure as seen on frequency-time spectra.  The cyclotron maser
instability (CMI) \citep{w79} is widely assumed to be the basic
plasma mechanism responsible for the emission. This mechanism originally
assumed a loss-cone electron velocity distribution function, but in situ
observations in the acceleration region have shown that a 'horseshoe' or
crescent distribution is more accurate \citep{l90,r93,d98,e00}. The
horseshoe distribution, which arises naturally for electron beams in the
presence of inhomogeneous magnetic fields \citep{s05}, provides a robust
and efficient free energy source for the CMI mechanism, as shown both by
model calculations \citep[e.g.][]{p84,p99,b00} and in laboratory experiments
\citep[e.g.,][]{s05}. This mechanism has been applied not only to
terrestrial AKR emission, but to many astrophysical environments in which
the requisite conditions (low density, beamed electrons, inhomogeneous
magnetic fields) are thought to be present, e.g. planetary magnetospheres
\citep{z98,f04}, stellar magnetospheres \citep{l86,k02,w04}, and even
relativistic jets in active galaxies \citep{b05}. 

Detailed physical models for the rich variety of observed spectral signatures
are still an active area of research. AKR fine structure
exhibits many morphologies on time-frequency spectra, including slowing
drifting or nearly stationary features lasting tens of seconds
\citep{g79}, drifting features which may be interacting with each other
\citep{ptb01}, and periodically modulated or banded emission \citep{g82}.
Although some AKR bursts appear to be broadband emission extending over
100's of kilohertz \citep{h01}, the total power is dominated by highly
time-variable fine structures, especially during periods of enhanced
bursts \citep{mm02}, suggesting that perhaps all AKR radiation is a
superposition of narrowband, short duration fine structures.

\citet{g79} first suggested that drifting AKR fine structure may be due
to localized sources
rising and falling along auroral magnetic field
lines,
with emitted frequency equal to the local electron cyclotron frequency.
\citet{ga81} suggested that the emission is triggered by electrostatic
waves which drift along the field at the local ion-acoustic speed. An
early detailed model which attempts to explain AKR fine structure is the
tuned cavity model of \citet{c82} in which the source region acts as a
waveguide with sharp density boundaries.  The source emits radiation in
normal modes analogous to an optical laser, resulting in narrowband
emission which drifts as the wave packet propagates into regions of
varying width. A related idea is that of \citet{f95} in which the cavity
boundary is oscillating quasi-periodically.  Under these conditions,
broadband radio photons will be stochastically accelerated, resulting in
quasi-monochromatic discrete tones, whose frequencies slowly drift with
changes in the cavity geometry.  A difficulty with these models is that
they require special conditions (e.g. sharp density boundaries,
oscillating walls) which are not supported by observations \citep{p02}.
\begin{figure*}[t]
\centerline{\includegraphics[width=6in]
{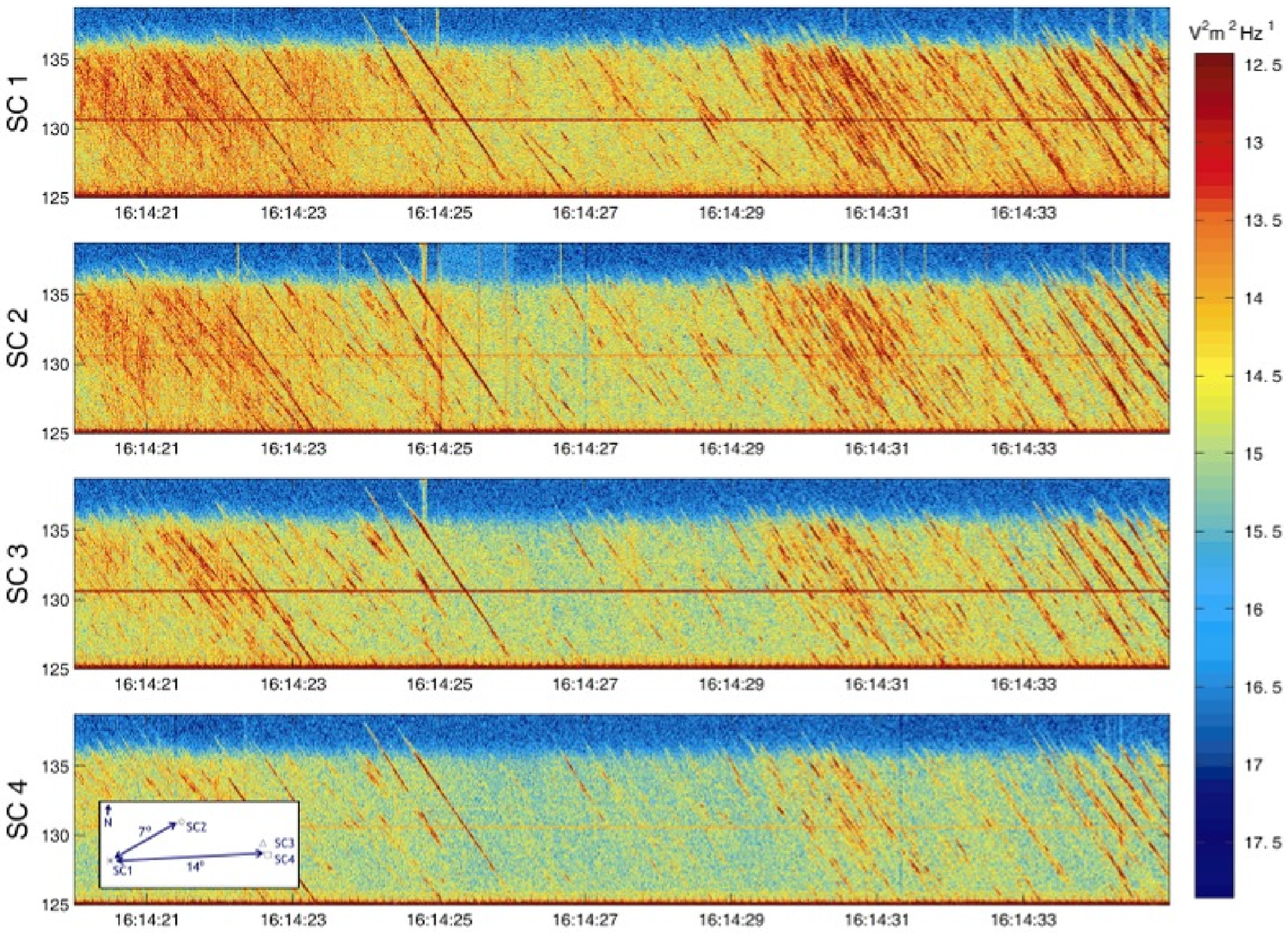}} 
\caption{Frequency-time spectra
of striated AKR bursts observed on 4 Cluster spacecraft on  August 31,
2002 from 16:14:20 - 16:14:35 UT. The inset shows the projected angular
separations of the spacecraft as seen from the source.}
\label{fig-dynspec1} 
\end{figure*}

Several authors have reproduced AKR fine structure using electromagnetic
particle simulations to model the AKR source region. \citet{mw91} used a
one-dimensional electromagnetic particle-in-cell (PIC) code to model the
cyclotron maser instability in a plasma with an inhomogeneous magnetic
field. They found that radiation is emitted in individual packets which
combine to form drifting features that both rise and fall, consistent
with some types of AKR fine structure. Both X- and O-mode drifting fine
structures are created, consistent with observations \citep{b88} but not
predicted in feedback models such as \citet{c82}. A weakness of their
model is that predicted drift rates are generally much higher than
observed drifts.  \citet{p02} used a 2-dimensional PIC code with model
parameters based on in situ FAST spacecraft measurements of the AKR
source region.  The simulated AKR is most strongly amplified  
longitudinally and consists entirely of short-timescale fine structures.
The predicted bandwidth ($\Delta\omega/\omega\sim10^{-3}$)
is somewhat larger than at least some observed AKR fine structure.   

More recently, Pottelette and colleagues have
investigated the possible connection between AKR fine structure and
electron holes \citep{ptb01,ptbj03} and also tri-polar structures
\citep{pt05}. AKR radiation from small packets  ('elementary radiation
sources') associated with the holes are thought to be strongly amplified
as the packets slow down and are reflected from the field-aligned
electric field.  Since the scale size of electron holes is very small (a few
Debye lengths, \citet{b99}), the observed narrow bandwidth of AKR fine
structures is easily accounted for. However, since neither the observed
speeds or direction of
electron holes (1000 - 2500 km s$^{-1}$ always downward, \citep{b99}) are
representative of fine structure frequency drifts, there is not a
straightforward
relationship between the observed properties of electron holes 
and the dynamics of the AKR fine structure.

In this paper, we present evidence that a particular type of AKR fine
structure called striped or striated AKR \citep{m96,mppg00} is triggered
by ion solitary structures (ion holes). In section 2 we present new
observations of SAKR bursts, including new bandwidth and angular
beamwidth measurements.  In section 3 we calculate the change in CMI gain
for a density depleted region with a pre-existing horseshoe velocity
distribution when perturbed by a passing ion hole. We compare the
observed properties of SAKR with those calculated by radiation
generated from the CMI instability in an ion hole. In section 4 we
discuss some inferred properties of ion holes, such as lifetimes,
probability of occurrence as a function of location, and radial dependence
of speed, that cannot be inferred from single spacecraft in situ measurements.

\section{Observed properties of SAKR}

Striated AKR (SAKR) consists of trains of narrowband drifting bursts with
negative slopes in the range -2 to -20 kHz s$^{-1}$, corresponding to
upward-traveling sources with speeds between 100 - 1,000 km$s^{-1}$.
Figure \ref{fig-dynspec1} shows a typical dynamical spectrum of SAKR bursts
observed with the Cluster wideband (WBD) plasma wave instrument \citep{g97} on
the four Cluster spacecraft. The bursts were observed in the 125-135 kHz
band between 16:14:20 - 16:14:35 UT on 31 August 2002. Although the
spectra are similar on all four spacecraft, there are clear differences
in individual striations, indicative of angular beaming on the scale of
the projected spacecraft separations, as shown in the inset at lower
left. Also, individual bursts have very narrow bandwidths and are nearly
always observed in groups with spacing between bursts of 30-300 ms. These
characteristics are described in more detail below.  

\subsection{Occurrence frequency}
The occurrence probability  of SAKR emission is quite low at
frequencies above 100 kHz: We
detect SAKR bursts in less than 1{\%} of all WBD spectra observed when
the spacecraft was above 30\deg magnetic latitude (Note that below this
latitude, there is often shadowing by the Earth's plasmasphere.) The
occurrence probabilities were computed by dividing the number of dynamic
spectra (length 52 sec) for which
SAKR emission was clearly detected by the total number of
spectra over a one year interval (July 2002 to 2003). 
This overestimates the actual occurrence probability for SAKR since we do
not correct for the fractional time within each spectrum that SAKR is
present.

\begin{figure}
\centerline{\includegraphics[angle=-90, width=\hsize]
{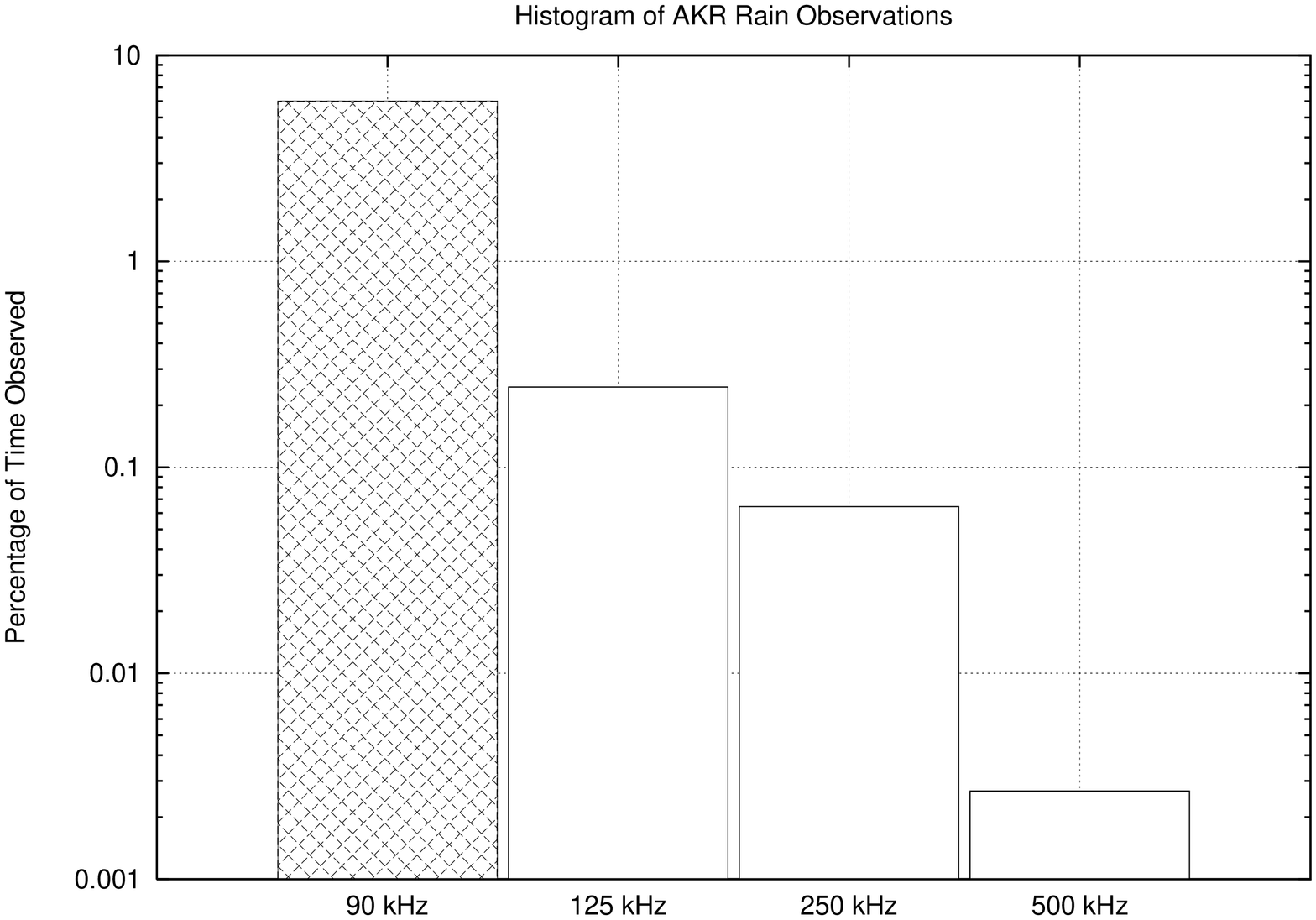}} 
\caption{Frequency of occurrence of SAKR as a function of observing
frequency.  The data for 90 kHz (hatched bar) is from \citet{mppg00}.}
\label{fig-histo-freq} \end{figure}

There
is a strong inverse correlation with observing frequency as shown Figure
\ref{fig-histo-freq}. There are only few detections in several hundred
hours of data in the 500-510 kHz band (detection probability $p<0.01\%$
), while $p\sim0.2$\% in the 125-125 kHz band.  \citet{m96,mppg00} also
studied SAKR emission using the plasma wave instrument (PWI) receiver on
the Polar spacecraft. They also found an inverse correlation with
observing frequency, with a 6{\%} detection rate in the 90 kHz band
(Fig. \ref{fig-histo-freq}, shaded bar) and comparable probabilities
and overlapping frequencies.  By contrast, non-striated AKR emission is
frequently detected, especially during sub-storm onset. The occurrence
probability of AKR is nearly 30{\%} for geomagnetic index K$_{p}$ in the
range 1$<$K$_{p}<$ 3, with the highest occurrence frequency in the winter
polar regions \citep{k98}.

\subsection{Frequency drift and speed}
\begin{figure}
\centerline{\includegraphics[angle=-90,width=\hsize]
{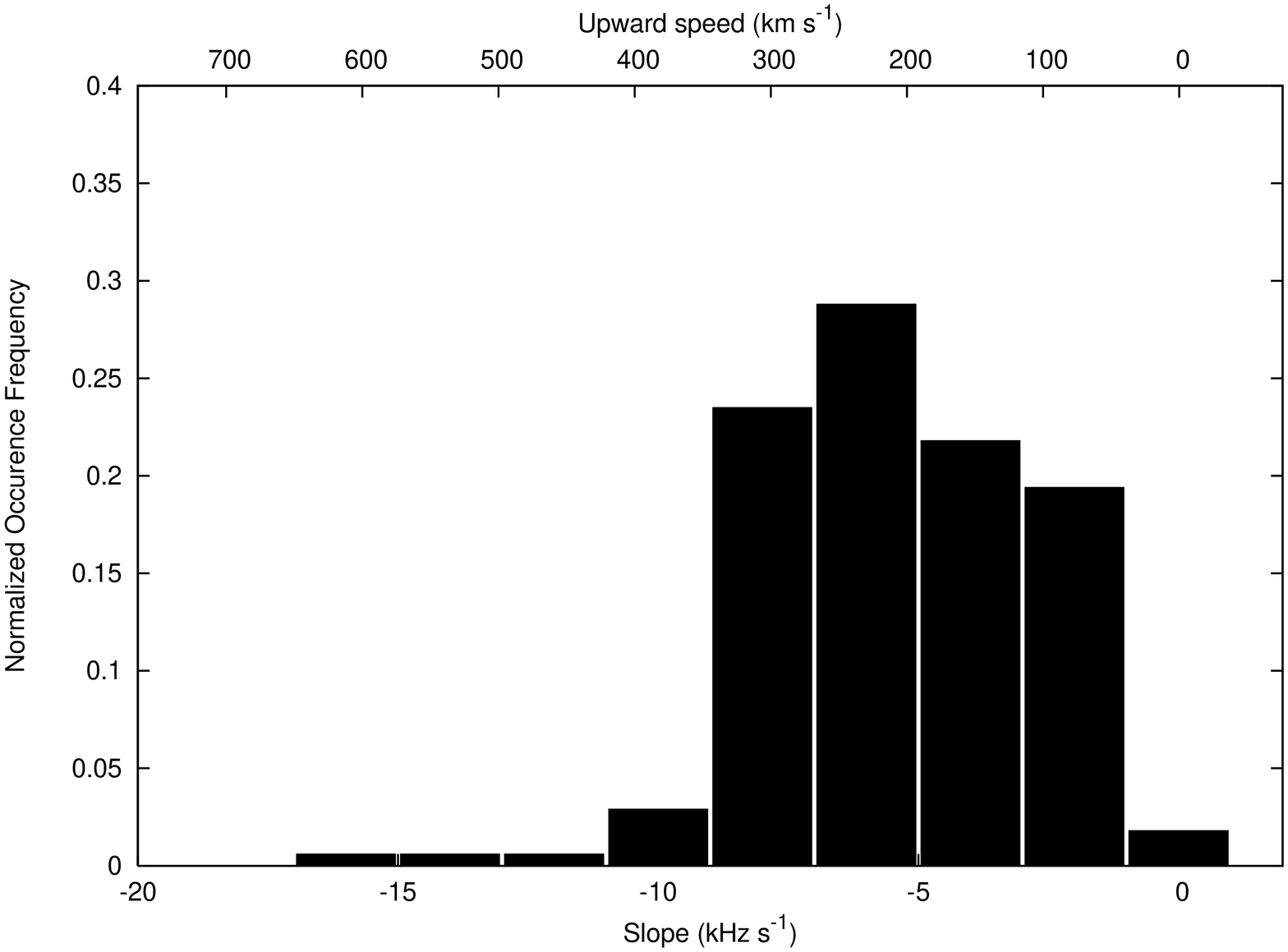}} 
\caption{Histogram of observed
SAKR burst slopes in the 125-135 kHz band (bottom x-axis) and derived
trigger speed (top x-axis, from equation \ref{eq-speed}).}
\label{fig-histo-slope} \end{figure}

SAKR bursts have negatively sloped, nearly linear morphologies, at least 
within the normally sampled bandwidth (9.5 kHz) of the WBD receivers. The majority of the 
slopes range between -2 kHz s$^{-1}$ and -8 kHz s$^{-1}$. Assuming the 
emission frequency is identified with the local electron gyrofrequency, the 
speed of the stimulator as a function of frequency and slope can be written 
\begin{equation}
\label{eq-speed}
V=540\rm{km\ s}^{-1}\cdot \left( {\frac{\alpha }{10\,kHz{\kern 1pt}s^{-1}}} 
\right)\,\left( {\frac{100\,kHz}{f}} \right)^{\frac{4}{3}}
\end{equation}

where V is the stimulator speed (km$s^{-1})$, \textit{$\alpha $} is the
observed slope of the SAKR burst (kHz s$^{-1})$, $f$ is the observed
frequency (kHz), and we have assumed the source moves upward along a dipolar
magnetic field line at magnetic latitude $\lambda_M$ = 70\deg. Figure \ref{fig-histo-slope} shows a histogram of the observed slopes of 650
SAKR burst events observed in the 125-135 kHz band along with the derived
trigger speeds (top x axis).  The mean slope is -5.6 kHz s$^{-1}$, with
more than 90{\%} of the slopes in the range -2 to -8 kHz s$^{-1}$. The
corresponding trigger speeds, using equation \ref{eq-speed}, are in the range 76 --
303 km s$^{-1}$, with a mean value 213 km s$^{-1}$. These results are
similar to those reported by \citet{mppg00}.

\begin{figure*}
\centerline{\includegraphics[width=6in]
{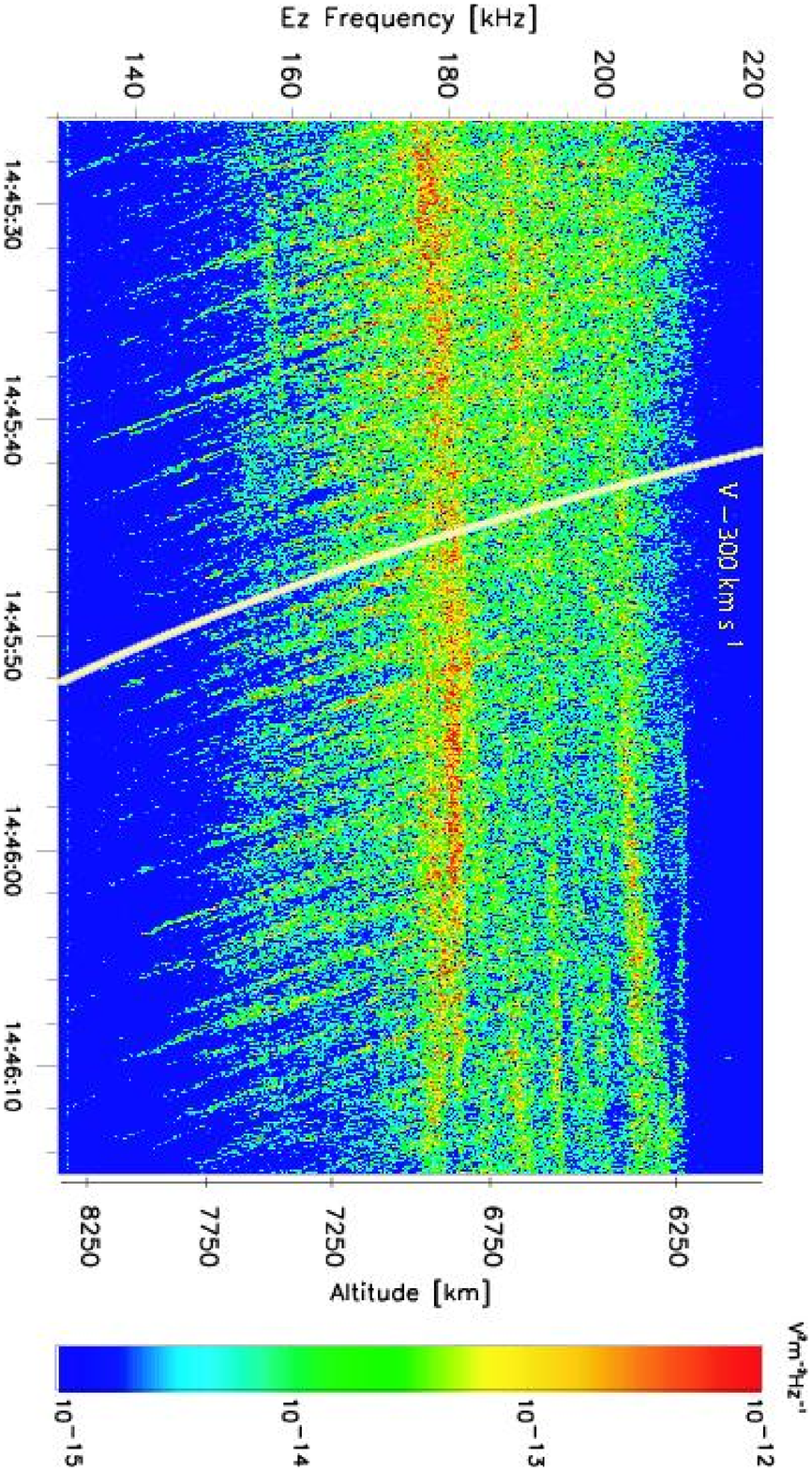}}
\caption{Striated AKR bursts observed on Cluster spacecraft SC2 on Jan
15, 2005 from 14:45:26 to 14:46:15 UT. Individual burst trails extend
from 180 kHz to 135 kHz, corresponding to an altitude range  from 7,000
km to 8,100 km. The yellow line is a calculated trail for a constant
velocity source moving upward at 300 km s$^{-1}$. Note that the SAKR
bursts appear to originate at or near an intense, slowly drifting AKR
source at $\sim7000$ km.} 
\label{fig-dyn-spect2} \end{figure*}

Figure \ref{fig-dyn-spect2} shows a group of SAKR bursts detected on
one spacecraft while operating the WBD instrument in a  wider
bandwidth (77 kHz) mode. Over this wider bandwidth, the striations
are not linear, but rather have a curved, frequency-dependent slope. The
curve can be fit by assuming a source radiating at the local electron
cyclotron frequency and moving with constant speed along a field line in
a dipolar magnetic field.  The  overlaid white line shows the expected
trace for a source moving upward at a constant velocity of
300 km$s^{-1}$. SAKR bursts with similar frequency-dependent behavior are
seen in Plate 1 of \citep{mppg00}. 

The altitude range of the SAKR locations are shown on the right y-axis.
For this example, individual bursts originated near 6500 km altitude and
moved upward at 300 km$s^{-1}$ to an altitude near 8100 km, implying a
lifetime of several seconds.  We have examined several other SAKR events
with wide bandwidth and have found that all exhibit similar lifetimes and
have nearly constant speed, although the speed varies with epoch.  

\subsection{Coeval broadband AKR emission}
SAKR bursts are almost always detected superposed on broader band AKR
emission.
This
can be clearly seen in both Figure \ref{fig-dynspec1} and
\ref{fig-dyn-spect2}, as well as in Plate 1 of \citet{mppg00}. This
appears to be a universal characteristic of SAKR, except in cases where the
SAKR intensity is so low that underlying broadband emission may have been
undetected. Background AKR emission coeval with SAKR bursts is consistent
with the hypothesis that ion holes trigger the SAKR
bursts (section 3.2) since the CMI gain in the source
region is significant even in the absence of the ion holes. 

\subsection{Bandwidth}
\begin{figure}
\centerline{\includegraphics[width=\hsize]
{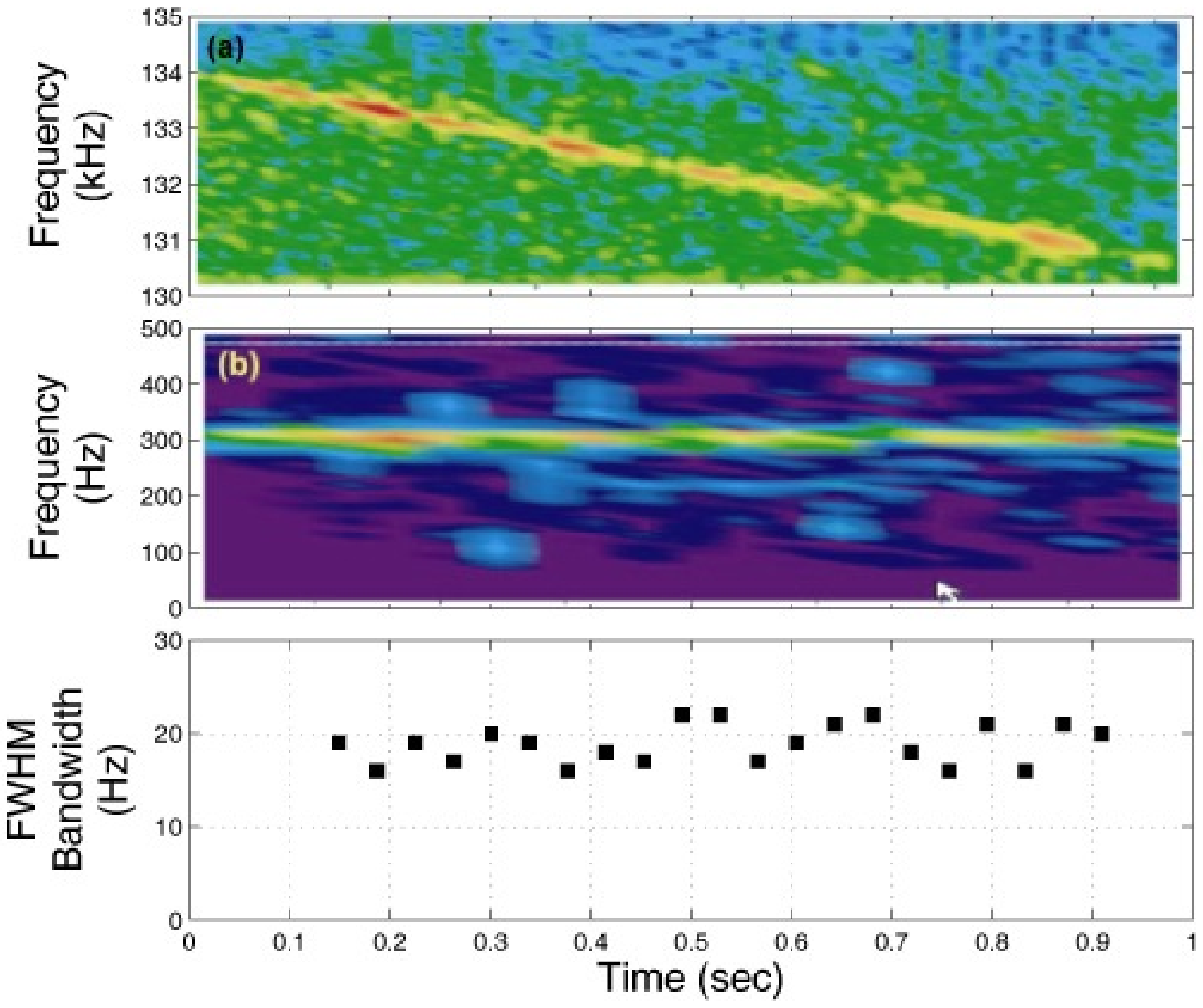}}
\caption{SAKR bandwidth. (a) SAKR burst observed with Cluster WBD, (b) Same 
burst, but detrended, (c) Measured bandwidth in 37 ms intervals.} 
\label{fig-bw} 
\end{figure}

One of the most unusual features of SAKR bursts is their extraordinarily 
narrow bandwidth. The top panel of Figure \ref{fig-bw} shows a single isolated SAKR 
burst. The middle panel shows the same feature, but after de-trending by 
subtraction of a `chirp' signal of best-fit constant negative slope. The 
lower panel shows the full width at half maximum of a Gaussian fit 
along the frequency axis of the de-trended signal after summing in time 
intervals of 37 ms each. The resulting bandwidths, in the range 15 -- 22 Hz, 
are much narrower than previously reported either observationally (e.g. 
\citealt*{g79}) or resulting from model calculation of AKR 
emission (e.g. \citealt*{p99,y98}) 
although \citet{bc87} also reported AKR 
bandwidths as small as 5 Hz. (In the latter paper, the bursts do not 
appear to be SAKR emission). Other SAKR bursts we have examined have 
bandwidths ranging from 15 to 40 Hz.

The narrow bandwidth implies a small source extent along the z (B-field) 
direction, viz, 
\begin{equation}
\label{eq-deltaz}
\Delta z=0.55\,km\cdot \left[ {\frac{100\,kHz}{f}} 
\right]^{\frac{4}{3}}\left[ {\frac{\Delta f}{10~Hz}} \right]
\end{equation}

where $\Delta $z is the source extent along the B field and $\Delta $f is the 
observed bandwidth. For $f = 130$ kHz and 
$\Delta $f = 20 Hz we obtain $\Delta $z = 0.76 km. This is considerably 
smaller than the lateral extent of the AKR source region and indicates the 
trigger for SAKR must have a dimension along the magnetic field of order 1 km.

\subsection{Angular Beam Size}
The Cluster spacecraft array has the unique ability to 
simultaneously sample the flux density of individual AKR bursts at four 
widely separated points in space. By comparing the flux density on pairs of 
spacecraft, we can estimate the average angular beam size of individual 
bursts. The beaming probability is defined using the following algorithm. 
For each pair of spacecraft during a given observation, we calculate the 
projected angular separation between spacecraft as seen from a location 
situated above the magnetic pole of the hemisphere being observed, at a 
height corresponding to the electron cyclotron frequency at the 
center frequency of each observing band (e.g., 2.35 R$_{e}$ for the 125-135 kHz band). This 
constitutes an average AKR location in that band without regard to location 
on the auroral oval. We next correct for differential propagation delay
by shifting the waveform data from each spacecraft to the distance
of the nearest spacecraft. We then divide the time-frequency
spectrum from each spacecraft pair (52 sec duration for 125-135 kHz band)
into data `cells' 19 msec x 53 Hz in size. This window was chosen to
match the observed bandwidth of SAKR (cf. section 2.4) but the beaming
results are not very sensitive to the data window size. We computed the
angular beam size with data windows factors of two smaller and larger, and
the results did not differ significantly.

The intensities are normalized
to correct for differing distances between individual spacecraft and the
AKR source. We then omit from further analysis all data cells whose
intensities are below a threshold, arbitrarily chosen to be 10 dB below
the maximum flux density for that spectrum. Finally, we compare
intensities in each pair of data windows, assigning a weight 1 to pairs
for which the intensities are within 10 dB of each other, and 0
otherwise. The overall beaming probability for each angular separation
interval is the sum of the beaming weights divided by all cell pairs. 

Note that this scheme is susceptible to overestimation of the beaming 
probability, since it is possible that independent AKR sources will 
illuminate separate spacecraft at the same frequency and time interval. This 
`confusion' problem is smaller for SAKR emission since it is often clear 
from the burst morphology on a time-frequency spectrum that only one source 
contributes to a given data cell at one time, whereas with normal AKR there 
are very often several intersecting sources which contribute to a 
given data cell.

\begin{figure}
\centerline{\includegraphics[width=\hsize]
{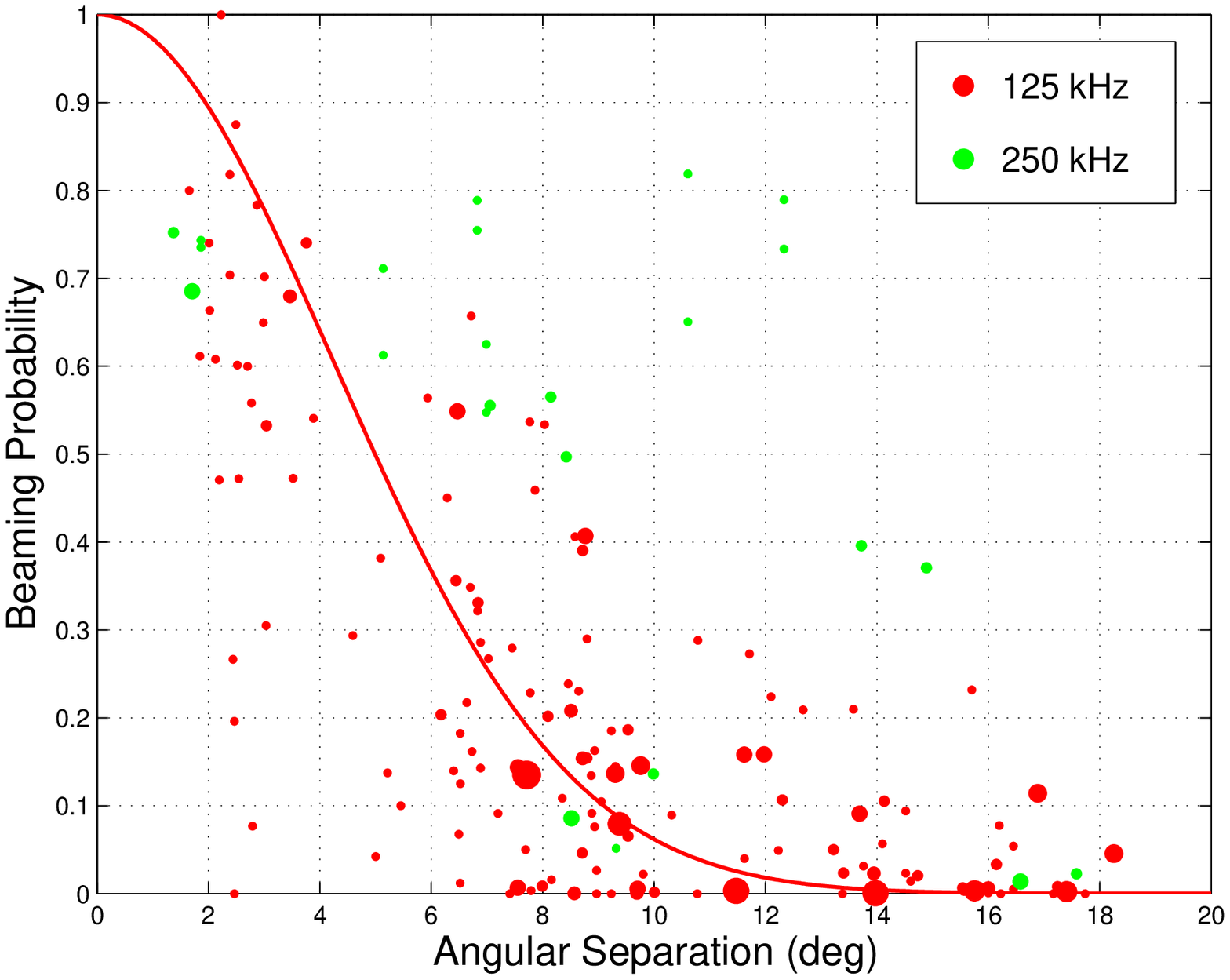}}
\caption{Angular beamsize for 651 individual SAKR burst sources 
measured at 125 kHz (red dots) and 250 kHz (green dots) using observed 
correlations between multiple spacecraft. The size of each point is
proportional the number of bursts for that point. The fitted FWHM beamsize is
5.0\deg.} 
\label{fig-angbeam} 
\end{figure}

In Figure \ref{fig-angbeam} we plot the beaming probability versus angular
separation for 651 SAKR bursts in the 125-135 kHz and 250-260 kHz bands.
We have fitted a Gaussian function to the 125 kHz band observations using
a least-squares fitting algorithm. The resulting full width at half
maximum (FWHM) angular size is \textit{$\theta $} = 5.0\deg\ (solid angle
$\Omega $ = 0.006 sr). This is surprisingly small compared with most
previously published observations of AKR beam size (e.g. \citet*{gg85}),
who reported beaming solid angles of 4.6 sr and 3.3 sr at 178 and 100 KHz
respectively. They made angular beaming estimates by comparing
time-averaged spectra observed using two satellites (Hawkeye and IMP-6)
which simultaneously observed AKR bursts while the spacecraft were both
over the same polar region. Since the time resolution used for the
Hawkeye-IMP6 spectrum comparison (several minutes) far exceeds the
time-scale of individual AKR bursts, their measured angular beam is
actually a measure of the ensemble-averaged sky distribution of AKR
bursts over a several minute time-scale rather than the angular beam size
of individual AKR emission sources. This is the confusion problem
mentioned above. 

An important unanswered question is the 2-dimensional structure of the SAKR
burst angular
beam pattern: It is asymmetric, or perhaps a hollow cone as suggested by
\citet{c87}? Since
the spatial frequency coverage of the Cluster spacecraft array is often
nearly one-dimensional at high magnetic latitude, it is difficult to
analyze the 2-dimensional structure of individual bursts. The angular
beaming probability plot shown in Figure \ref{fig-angbeam} includes
measurements over a range of baseline orientations. However, a preliminary
analysis of beaming probability grouped by baseline orientation did not
reveal any obvious trends. We are
presently analyzing a much a larger dataset consisting of a large variety
of AKR emission and wider range of baseline orientations 
to investigate the 2-dimensional structure of the AKR
angular beam.

\subsection{Individual SAKR Source Power Estimate}
The angular beaming observations, combined with measured flux density,  allow a direct estimate of the average
intrinsic power of individual SAKR bursts. For intense SAKR bursts, 
the observed square of the electric field
intensity is \begin{equation} I_{\nu}\sim (1-10)\times 10^{-12} {\rm\
(V/m)}^{2}{\rm\ Hz}^{-1} \end{equation}

at a source-spacecraft distance $d$ = 10R$_{e}$. Converting to
flux density, we obtain
\begin{equation}
S_{\nu} = I_{\nu}/Z_0 \sim (0.3-3)\times 10^{-14}\ {\rm\ W\ m}^{-2}{\rm\ Hz}^{-1}
\end{equation}
where $Z_{0}$ = 377 ohms is the impedance of free space. This flux
density range is about one hundred times smaller than the flux density of
intense AKR bursts reported
by \citet{bf91}. This is consistent with the conjecture that intense AKR bursts are the sum of
many spatially distinct 'elementary radiation sources'  \citep{ptb01}

The power emitted at
the AKR source is the isotropic power corrected by the angular beamsize
of an individual SAKR burst 
\begin{equation} \label{eq-pwr} P=4\pi
d^2\cdot S_\nu \cdot \Delta \nu \cdot \frac{\Omega }{4\pi }
\end{equation}

where \textit{$\Delta \nu $} $\sim $ 20-50 Hz is 
the bandwidth of a burst, and \textit{$\Omega $} $\sim $ 0.006 sr, is the solid angle of the 
emission beam. Using these values,  the resulting power is P $\sim $ 1 -- 10 W, much smaller than 
the previous estimates of P $\sim $ 10$^{3}$-10$^{4}$ W for a single 'elementary 
radiator' \citep{ptb01}.

\section{Connection with ion holes}
Ion solitary structures, also known as ion holes, are small-scale
($\sim$1km) regions of negative electrostatic potential associated with
upgoing ion beams. They are seen in spacecraft electric field
measurements as symmetric bipolar parallel electric field structures with
amplitudes 10 -- 500 mV m$^{-1}$ and timescales 3-10 ms. They were first
detected in S3-3 spacecraft observations \citep{t82} and have been
subsequently been studied using in situ measurements in the acceleration region by several
authors (e.g.  \citet*{b99,d01,m03}). Ion holes travel upward at 
speeds between 75-300 km$s^{-1}$ \citep{b99,d01} (although \citet{m03}
argue for somewhat higher speeds, in the range 550 - 1100 km$s^{-1}$). The
width of the waves increase with amplitude \citep{d01} which is
inconsistent with small amplitude 1-d soliton models, but which supports
a BGK-type generation mode \citep{m02}. 

\subsection{Electron distribution function and CMI growth rate}
AKR radiation arises from wave growth resulting from the 
interaction of a radiation field with an electron velocity distribution 
having a positive slope in the direction perpendicular to the magnetic field 
($\partial f$/$\partial $v$_{\bot }>$ 0). The condition for waves of 
angular frequency $\omega $ and wave normal angle $\theta $ to resonate with 
electrons with angular frequency $\omega _{ce}$ and velocity $v$ is given by 
\citep{w79}
\begin{equation}
\label{eq-cmi1}
k_\parallel v cos\,\theta -\omega +n\frac{\omega _{ce} }{\gamma \left( 
{v_\parallel ,v_\bot } \right)}=0
\end{equation}

where $\omega _{ce }$ is the electron cyclotron frequency, $n$ is an integer, 
and $\gamma $ is the Lorentz factor of the relativistic electrons. The 
cyclotron maser instability (CMI) is the case $n $=1. If we assume that the 
electrons are mildly relativistic, as is observed in the AKR source region 
(E$_{e }\sim $ 10 keV, so $\gamma\sim$1.01), we can expand the 
Lorentz factor to obtain the equation for a circle in velocity space 
\begin{equation}
\label{eq-cmi2}
v_\bot ^2+\left( {v_\parallel -v_c } \right)^2=v_r ^2
\end{equation}

where the resonant circle's center is on the horizontal axis displaced by
\begin{equation}
\label{eq-cmi3}
v_{c} ={\frac{k_\parallel }{\omega _{ce}}\ c^2 }
\end{equation}

and the radius of the resonant circle is
\begin{equation}
\label{eq-cmi4}
v_r =v_c \left[ {1-\frac{2c^2\left( {\omega -\omega _{ce} } 
\right)}{{v_c}^2\omega _{ce} }} \right]^{\frac{1}{2}}
\end{equation}

Analysis of recent FAST observations of AKR emission in the source region 
\citep{e98,e00,p99} 
provides evidence, based on wave polarization, that the AKR k-vector direction at the source is nearly 
perpendicular to the B field ($k_{\parallel}\sim0$). The E-field of 
the wave is polarized perpendicular to the ambient magnetic field which 
indicates that the wave is purely X-mode \citep{s01}. In 
the following, we explicitly assume $k_{\parallel}=0$, so that the 
radius of the resonance circle becomes
\begin{equation}
\label{eq-cmi5}
v_r =c\left( {\frac{-2\delta \omega }{\omega _{ce} }} \right)^{\frac{1}{2}}
\end{equation}

where $\delta \omega =\omega$ -- $\omega_{ce} $. The growth rate of the CMI 
mechanism is given by calculating the imaginary part of the angular 
frequency,
\begin{equation}
\label{eq-cmi6}
\omega _i =\frac{\pi ^2\omega _{pe} ^2 }{4\omega n_e }
\oint{v_\bot ^2\frac{\partial f}{\partial v_\bot }dv_\bot}
\end{equation}

where the $v_{\bot }$ integral is performed on the closed circular path given 
by equation 4. 

\begin{figure}
\centerline{\includegraphics[width=\hsize]
{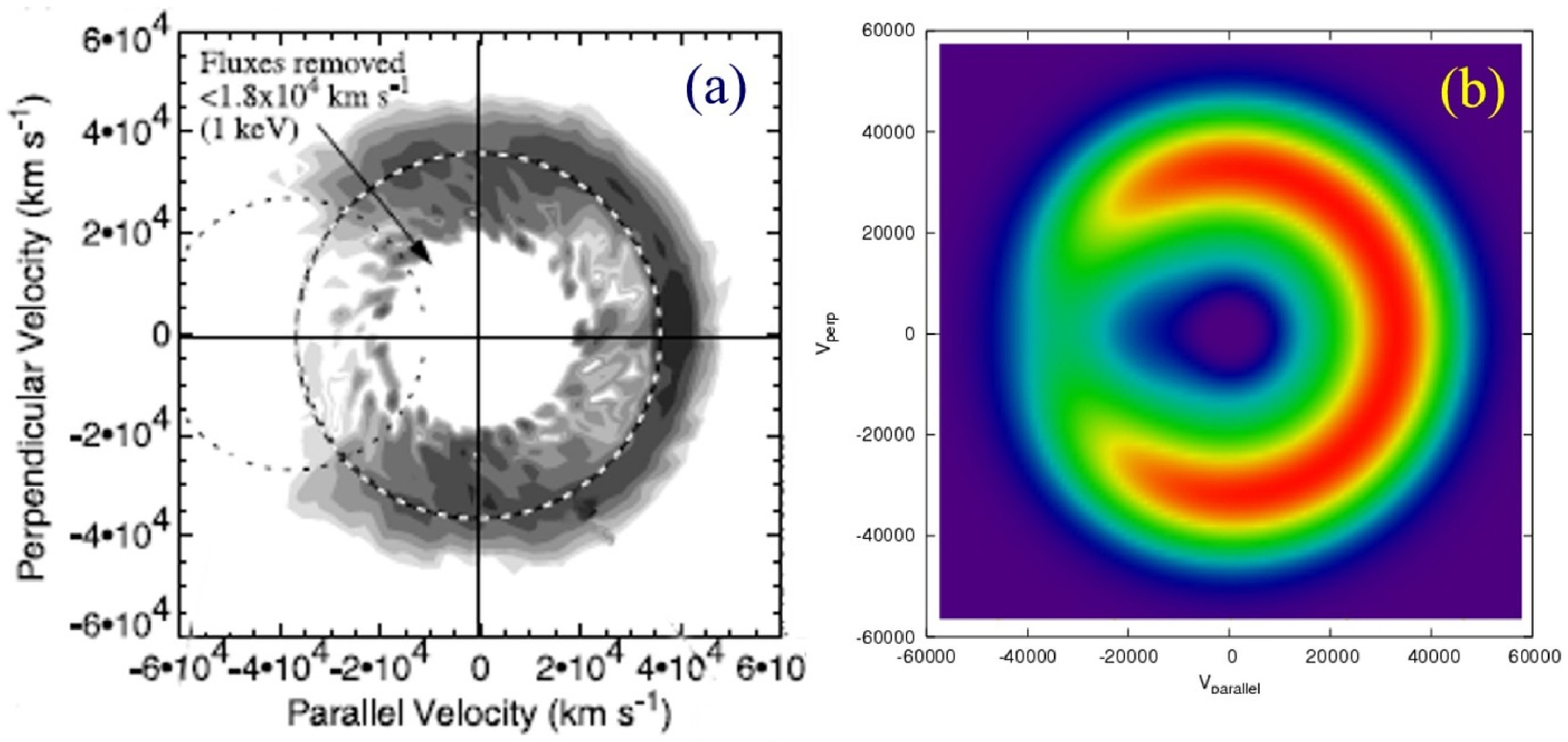}}
\caption{(a) Electron distribution function measured by FAST in AKR source 
region \citep{e00}. (b) Model 
distribution.} \label{fig-dist-2x1}
\end{figure}

The electron velocity distribution in the upward current acceleration
region has been measured \textit{in situ} by both Viking \citep{r93} and
FAST \citep{d98} spacecraft. It consists of an incomplete shell or
`horseshoe' shape in velocity space. The density of cold electrons (E
$<<$ 1 keV) in this region is much smaller than the hot electron population
which comprises the horseshoe component \citep{s98}. In this paper we
assume that all electrons are in the hot component.

We have modeled the observed
velocity distribution using a simple analytic functional form
\begin{equation} \label{eq-dist1} f(v) = g(v)\exp \left[ { - \left(
{\frac{{v - v_0 }}{\Delta }} \right)^2 } \right] \end{equation}
where \textit{$\Delta $} is the horseshoe width, $v_{0}$ the horseshoe radius and the loss-cone 
function $g(v)$
\begin{equation}
\label{eq-dist2}
g\left( v \right)=1-\beta \cdot \mbox{sech}\left( {\frac{\tan ^{-1}\left( 
{v_y /v_x } \right)}{\Theta }} \right)
\end{equation}
where $\beta $ is a dimensionless scaling factor and $\Theta $ is the 
characteristic opening angle of the loss cone. The model velocity distribution
function is shown in Figure \ref{fig-dist-2x1} along with a measured
velocity distribution function from FAST \citep{e00}.

\begin{figure}
\centerline{\includegraphics[width=\hsize]
{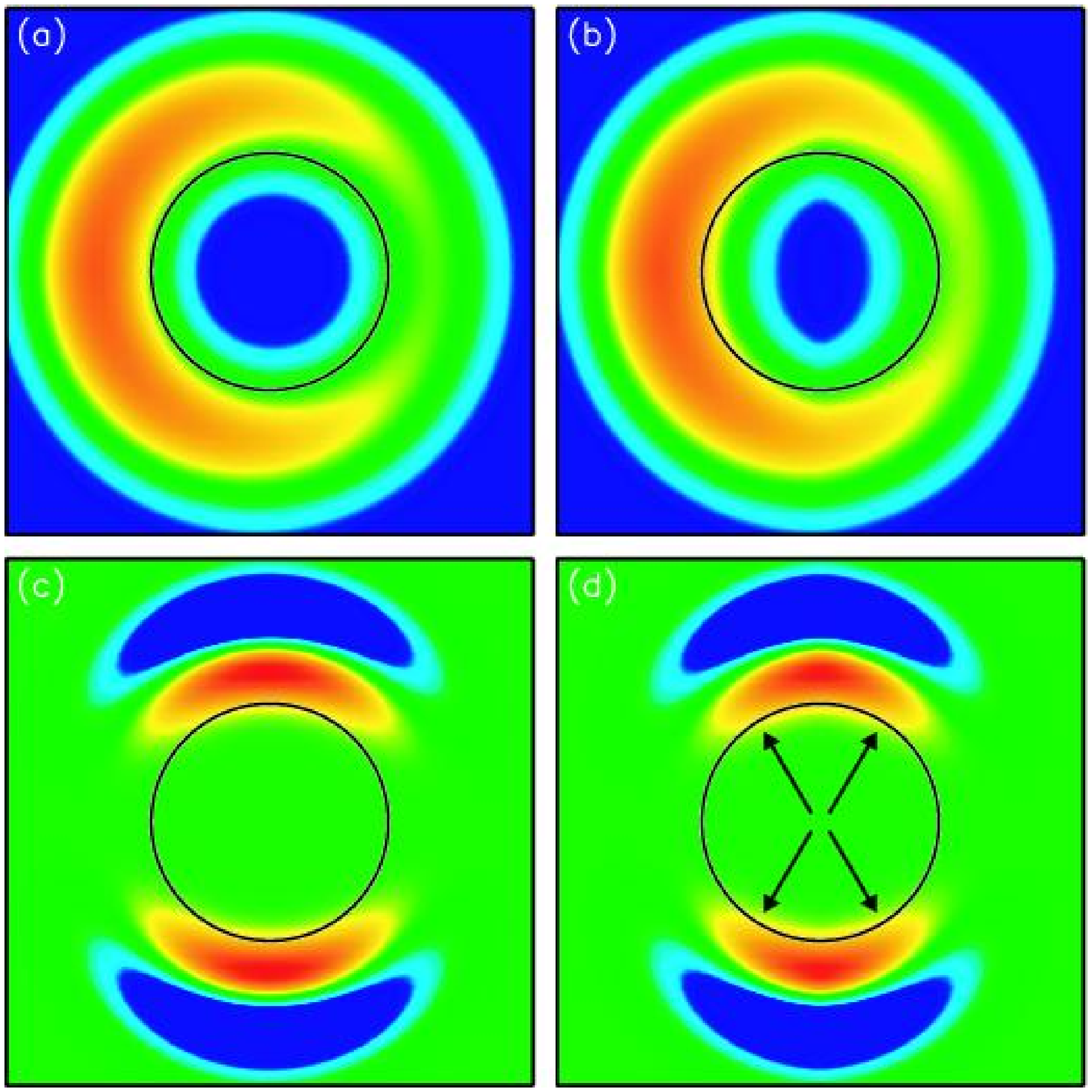}}
\caption{Top panel: horseshoe electron velocity distribution function outside 
an ion hole ($a)$, and inside an ion hole ($b)$. Black circle is the CMI resonant circle. The 
$v_{\vert \vert }$direction is the horizontal axis, Earthward to left. Bottom 
panel: weighted partial derivative $\partial f$/$\partial $v$_{\bot }$ 
outside an ion hole ($c)$, and inside an ion hole ($d)$. The arrows indicate the parts of the 
growth line integral (equation 9) enhanced inside the hole.}
\label{fig-elec-dist-4by1}
\end{figure}

\begin{figure}
\centerline{\includegraphics[width=\hsize]
{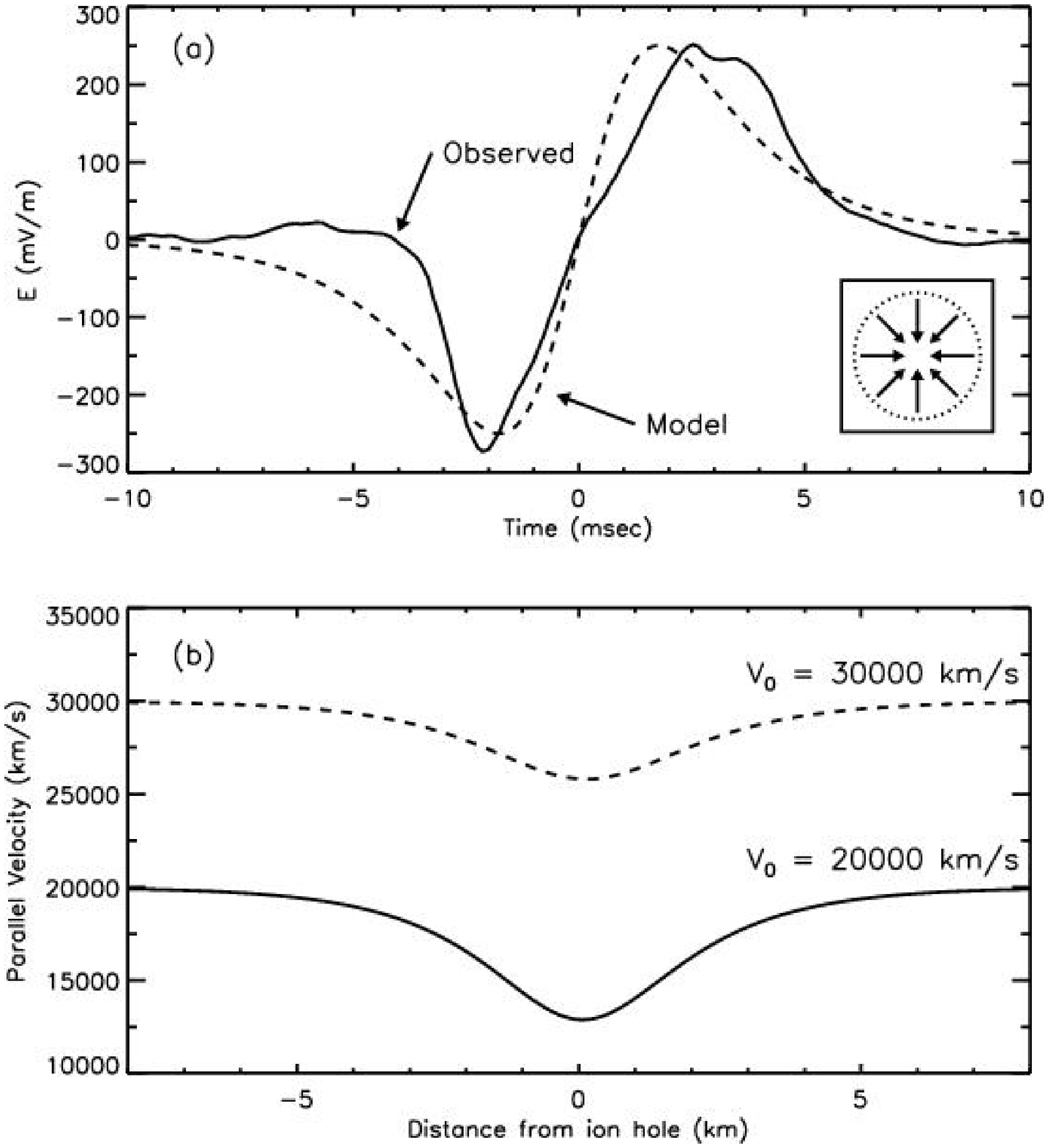}}
\caption{(a) Observed electric field vs. time (solid line) as an ion hole passes
by FAST spacecraft \citep{m03}, and best-fit model (dashed line)
using equation (\ref{eq-emod}). The inset at lower right shows a simple
spherical model of an ion hole with inward electric field. (b) Calculated
parallel velocity vs. distance from an ion hole for an electron passing through
an ion hole with an impact parameter 1 km and initial velocities 20,000 km
s$^{-1}$ (dotted line) and 30,000 km$s^{-1}$ (solid line). The assumed
hole speed is 300 km$s^{-1}$.} \label{fig-hole-efield} 
\end{figure}

\subsection{Triggering SAKR emission by an ion hole}
The range of possible interactions between background electrons and ion
holes is very complex and varied \citep{es05}. These include
trapping of electrons between ion holes, excitation and trapping of
high-frequency Langmuir waves from electron streams resulting from ion
hole collisions, and modification of
the ion holes via the ponderomotive force. In this paper we will tacitly
assume that the ion holes are well-spaced and non-interacting, and that
the only significant effect on the electron population is a transient speed
decrease as the electrons traverse the hole's negative potential well.
This assumption is motivated by the relatively simple structure of SAKR
bursts and by the success in using this assumption to explain all significant observed properties
of SAKR bursts. Other more complicated
forms of AKR fine structure seen on dynamical spectra may well involve
one or more of the complex interactions mentioned above.

Since the electrons are magnetized (gyro-radius
r$_{ce}\sim $50m, much smaller than the ion hole parallel scale size
r$_{ISS}\sim $1 km), the ion's hole's effect on the perpendicular
velocity component is negligible (a small \textbf{E}x\textbf{B} drift,
V$_{D} \quad \sim $ 40 m s$^{-1})$. However, the parallel velocity
component will
experience a significant decrease as the electrons are repelled by the
negative potential well of the hole. To determine the magnitude of the
effect, we have modeled the observed electric field structure of an
ion hole
using the analytic form \begin{equation} \label{eq-emod} E(t)=E_0 \tanh
\left( {\frac{t}{\tau }} \right)\rm{sech}\left({\frac{t}{\tau }} \right)
\end{equation} Figure \ref{fig-hole-efield}a shows an observed E-field of
an ion hole along with a plot of the model E-field with $E_{0}$ = 500 mV
m$^{-1}$, and $\tau = 2$ ms. 

We used this model to calculate the velocities of electrons as they traverse 
the hole for a variety of impact parameters and initial speeds. For initial 
parallel speeds of 20,000 and 30,000 km$s^{-1}$, the speed of 
an electron traversing an ion hole at a minimum distance of 1 km from the center 
is shown in Figure \ref{fig-hole-efield}b. As expected, the parallel
velocity briefly decreases, so that the horseshoe velocity distribution
is `squeezed' on the parallel (horizontal) axis, as shown in Figure
\ref{fig-elec-dist-4by1}b.
The speed decrease is proportional to $v^{-1}$ as expected by a simple
energy conservation argument: For an electric field E and ion hole diameter
L$_{ih}$, the potential well of the hole is approximately $\Phi \sim
$E$\cdot $L$_{ih}$. The electron loses kinetic energy as it traverses the
potential well, so that energy conservation requires 
\begin{equation}
\label{eq-energy1} 
\delta \left( {\frac{1}{2}mv_{\parallel}^2+q_e \Phi } \right)=0
\end{equation} 

Solving for the velocity change, 
\begin{equation}
\label{eq-energy2} 
\delta v_{\parallel} =\frac{q_e \Phi }{m_e v_0 } \end{equation} For an
ion hole electric field E $\sim $ 300 mV m$^{-1}$, diameter L$_{CMI}$ $\sim $
2km, and $v_{0}$ = 20,000 km$s^{-1}$ , the expected speed decrease is
$\delta v_{\parallel}\sim $ 5,000 km$s^{-1}$, in good agreement with the
exact calculation. 
\begin{figure*}
\centerline{\includegraphics[width=7in]
{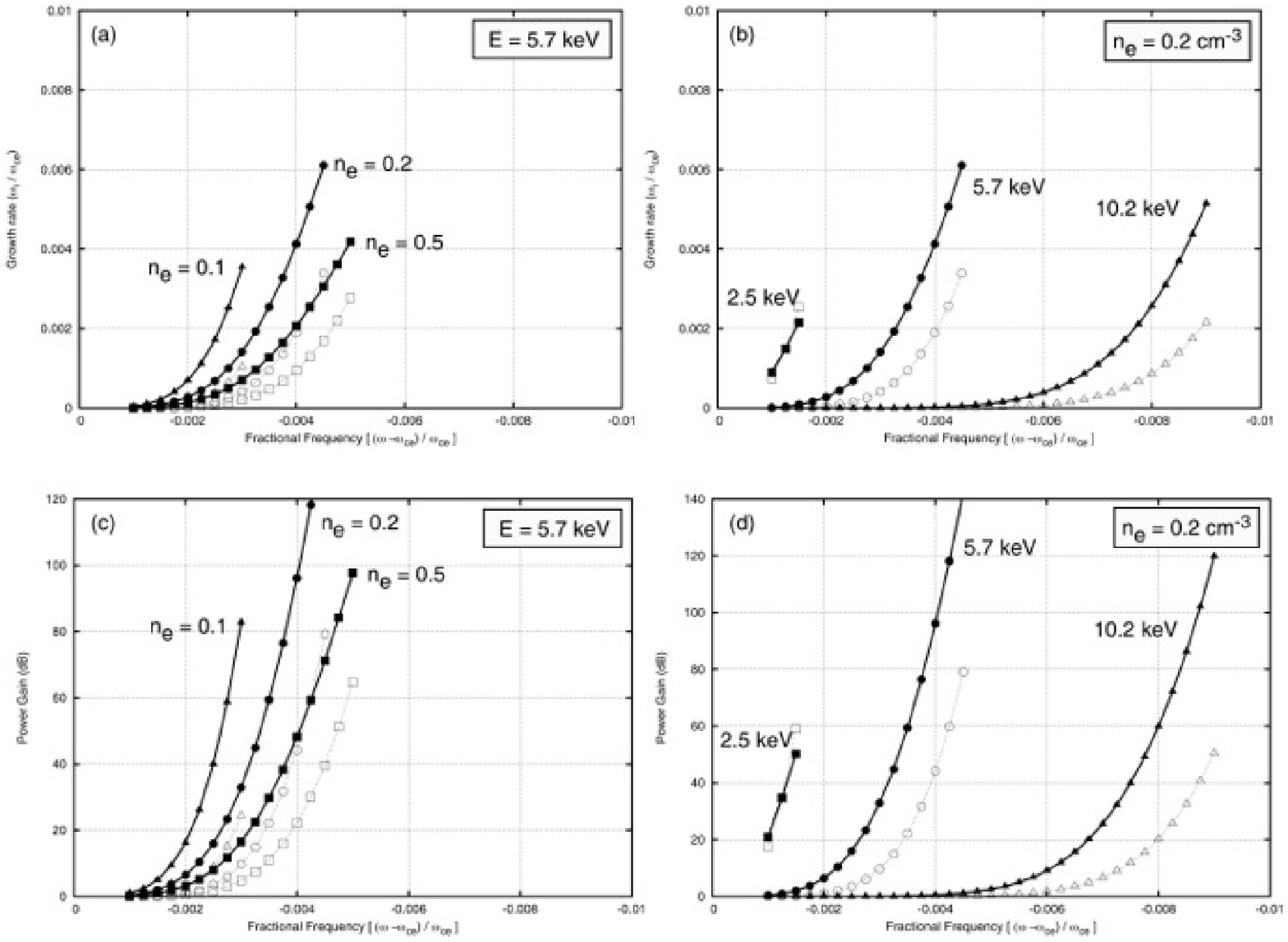}}
\caption{(a) Temporal growth rate as a function of fractional frequency for
electron horseshoe distribution radius $V_c/c =0.15$ ($\bar{E}$ = 5.7
keV) and electron density $n_e$ = 0.1 cm$^{-3}$($\triangle$), 0.2
cm$^{-3}$ ($\Box$), and 0.5 cm$^{-3}$
($\circ$). Solid lines are inside ion hole; dashed lines are
outside ion hole. (b). Same as (a), but with fixed $n_e$ = 0.2 cm$^{-3}$
and $V_c/c = 0.10$ (2.5 keV,$\Box$), $0.15$ (5.7 keV,$\circ$), $0.20$
(10.2 keV, $\triangle$).  (c)
CMI power gain vs. fractional frequency, same parameters as (a). (d)
CMI power gain vs. fractional frequency, same parameters as (c).
\label{fig-cmi-growth}}
\end{figure*}

In order to calculate the power gain in the perturbed region, we need an 
estimate of the convective growth length
\begin{equation}
\label{eq-lc}
L_c =\frac{V_g }{\vert \omega _i \vert }
\end{equation}
where $V_{g}$ is the group velocity of the wave and $\omega _{i }$ is the
growth rate. The group velocity is very sensitive to the ratio of plasma
to gyro-frequency and the detailed velocity distribution function. As a
first approximation, we have used the cold plasma dispersion relations to
estimate the group velocity $V_g = d\omega /dk$ for frequencies near the
R-mode cut-off frequency. We find that $Vg$ is of order 300 - 1000 km
s$^{-1}$, similar to the estimate used by \citet{og82}.

For a maximum growth rate $\omega _{i }\sim $ 5000 s$^{-1}$ (at
$\omega_{ce}$ = $2\pi\times125$ kHz), the
corresponding convective growth length is L$_{c}\sim$200m. Hence,
there are approximately 5-10 e-folding lengths in the region of the ISS,
assuming a perpendicular physical scale of order 1-2 km. This results in
a maximum power gain of e$^{10-20} \quad \sim $ 43-87 dB. The power gain
required to amplify the background radiation to observed levels of normal
(wideband) AKR was estimated by \citet{og82} to be e$^{20}$, which is at
the high end of our calculation. However, as discussed above, since SAKR
emission is highly beamed, the required power is a factor of $\sim $100 smaller,
so the requisite power gain is closer to e$^{15 }$= 65 dB. This is
comfortably within the range predicted by an ion hole trigger. 

Figure \ref{fig-cmi-growth} illustrates the growth rate and power gain
both inside and exterior to an ion hole for a range of relativistic
electron energies and plasma densities. Figure  \ref{fig-cmi-growth}$(a,
c)$ show the growth rate and power gain respectively for a horseshoe
velocity distribution with a radius $V/c = 0.15$ (E = 5.7 keV) and $n_e$ =
0.1, 0.2 and 0.5 cm$^{-3}$, while Figure \ref{fig-cmi-growth}$(b, d)$ shows
the growth rate and gain for $n_e$ = 0.2 cm$^{-3}$ and a range of radii
$V/c$ = 0.10, 0.15, and 0.20 (E = 2.5, 5.7, 10.2 keV). These values are
typical of the values of electron density and energy observed in the
auroral density cavity \citep{s98}. There are several noteworthy features
of these plots:
\begin{enumerate}
\item{The range of electron densities which
result in substantial CMI gain is very limited. The maximum density is constrained by the
condition that the R-mode cutoff frequency is less than the electron
cyclotron frequency. This can by expressed by the inequality
\begin{equation}
\frac{f_{pe}(n_e)}{f_{ce}} < \left[\frac{
\gamma -1}{\gamma}\right] ^{1/2}
\end{equation}
where $f_{pe}$ and $f_{ce}$ are the electron plasma and cyclotron
frequencies respectively, and $\gamma$ is the 
Lorentz factor. This inequality can also be written
\begin{equation}
n_e < {\left(
\frac{f_{ce}}{9\rm{kHz}}\right)}^2\left(\frac{E}{2E_0}\right)\ \rm{cm^{-3}}
\end{equation}
where $E_
0$ is the rest-mass energy of the electron (511 keV). For example, for  $E$
= 5 keV and $f$= 125 kHz, we find $n_e<0.9\ \rm{cm}^{-3}$. On the other
hand, for very low densities ($n_e << 0.1\ \rm{cm}^{-3}$), the CMI gain
decreases dramatically (cf. Fig. \ref{fig-cmi-growth}$a$)}

\item{The gain is a sharply peaked function of frequency. For
example, for $n_e$ = 0.3 cm$^{-3}$, $E=$ 5.7 keV
(Fig. \ref{fig-cmi-growth}$c$), a dynamic range of 30 dB near the peak
(approximately that observed on dynamic spectra of SAKR bursts) corresponds to
a fractional bandwidth $\delta f/f\sim$ 0.0005, or $\delta f\sim$ 50 Hz at
$f$ = 125 kHz, in good agreement with observed bandwidths (section 2.4).
These extremely narrow bandwidths are characteristic of CMI gain curves with
partial-ring velocity distributions \citep{y98}.}
\item{Even outside the ion hole there is substantial CMI gain under some
favorable conditions (e.g. 80 dB for the middle plot of
Figure \ref{fig-cmi-growth}$d)$.} This may be the explanation for the
background AKR seen in Figure \ref{fig-dyn-spect2} and discussed in section
2.3. 
\end{enumerate}

\subsection{Conversion efficiency}
Given the rather small volume of an ion hole, it is reasonable to ask if
there is sufficient free energy available from the resonant electrons to
power the observed SAKR emission. As shown in section 2.5, the observed
flux levels and angular beamwidths of SAKR emission correspond to
radiated powers  $P_{rad}\sim1-10$ W. Using a an ion hole scale
size $L_{ih}\sim$ 1 km and speed $V_{ih}\sim300\ \rm{km\ s}^{-1}$, and
mean electron density and energy $n_e = 0.5\ \rm{cm}^{-3}$, 
$E_e$ = 5 keV respectively, the ratio of radiated power to the
electron kinetic energy traversing a hole per unit time is
\begin{equation}
\frac{P_{rad}}{P_{tot}} =
\frac{(1-10) W}
{n_e E_e{L_{ih}}^2V_{ih}}\sim 0.002 - 0.02
\end{equation}

This range of conversion efficiencies is similar to previous models of
AKR emission \citep[e.g., ][]{p99} and indicates that while a single ion
hole extracts relatively little energy from the ambient electron
population, a train of hundreds of holes traversing the same region could
account for a significant modification of the electron distribution
function.

\section{Discussion}

If SAKR bursts are stimulated by upward traveling ion
holes in the acceleration region of the magnetosphere, then observations
of these bursts provides a new technique to study characteristics of ion
holes such as lifetimes and relative
number and speed as a function
of altitude, that are impossible to measure with single {\it
in situ} spacecraft .  
\begin{enumerate}
\item{The 77 kHz bandwidth observations of
SAKR bursts indicate that ion holes propagate upward for more than 1,000
km, implying lifetimes of a few seconds. This is much longer than
estimates of solitary wave lifetimes derived from PIC simulations of
solitary waves generated by the two-stream instability \citep{c01}, which
have lifetimes $\tau\sim{(100 - 1500)\ \omega_{pe}}^{-1}\sim$ 5 - 75 ms.}
\item{The majority of ion hole
speeds derived from SAKR observations are in the range 75 - 400 km
s$^{-1}$ (Fig. \ref{fig-histo-slope}).  These speeds are in very good
agreement with {\it in  situ} measurements of ion hole speeds from Polar
(75 - 300 km s$^{-1}$)  at altitudes between 5500 km - 7000 km
\citep{b99,d01}. However, \citet{m03} finds somewhat higher ion hole
speeds (550-1100 km s$^{-1}$) based on data from the FAST satellite at
altitudes between 3000 km - 4000 km.  It is possible that at speeds above
$V\sim\ $600 km s$^{-1}$ (slopes $>$20 kHz s$^{-1}$) SAKR bursts, especially in
closely spaced groups, would be  undetected since they would merge into
quasi-continuous emission on dynamic spectra. }
\item{Fits to SAKR bursts assuming a constant speed source, (Fig.
\ref{fig-dyn-spect2}) indicate that ion holes propagate at nearly
constant speed for their entire lifetime. Some numerical simulations
\citep{c01} show a significant change in ion hole speed as they evolve.
This is not supported by our observations.}
\item{SAKR bursts, and hence ion holes,  are much more
common at higher altitude, being more than one hundred times as common at
10,000 km ($f_{ce}\sim$ 90 kHz) than at 3,200 km ($f_{ce}\sim$ 500 kHz)
altitude, assuming ambient conditions favorable to the generation of SAKR
are not dissimilar in this altitude range.}
\item{Since SAKR bursts are almost always detected in  groups with typical
spacing 30 - 300 ms., 
(section 2.6), this likely also applies to ion holes. Both laboratory
plasma experiments \citep{f01} and observations of ion holes in the
magnetosphere \citep[e.g., ][]{b99} show trains of ion holes with
spacings similar to the SAKR bursts.}
\item{Finally, the
uniformity of SAKR intensity and bandwidth over a large frequency range
(Fig.\ref{fig-dyn-spect2}) implies that there is little evolution of the
electric field intensity or spatial structure of ion holes over their
lifetime.}
\end{enumerate}

\section{Summary and Conclusions}
This paper summarizes the observed properties of a distinct form of
auroral kilometric radiation fine structure called striated AKR, first
described by \citet{m96}. We present new observational results using the WBD
instrument on Cluster which characterize the bandwidth and angular
beamsize of individual SAKR sources, as well as derived properties such
as intrinsic power and speed along the magnetic field. 

Assuming the
frequency of SAKR bursts can be identified with the local
electron cyclotron frequency, the speed and direction of SAKR sources
calculated from their observed drift rates are very similar to those
observed for ion solitary structures (ion holes) in the upward current
region of the magnetosphere. Hence, we investigated whether SAKR bursts
could be the result of enhancement of the cyclotron maser instability by
ion holes. Using observed electric field signatures of ion holes in this
region, we calculate the perturbation caused by the passage of an ion
hole on a 'horseshoe' electron velocity distribution in a dilute
($n_e<<1$ cm$^{-3}$) plasma. The cyclotron maser instability is strongly
enhanced inside the ion hole, with power gain exceeding 100 dB in a
narrow frequency range just above the x-mode cutoff frequency. These
characteristics are in excellent agreement with the observed bandwidth, speed,
direction, and flux density of SAKR bursts. Alternative suggestions
involving other types of solitary structures to explain
AKR fine structure, such as electron holes \citep{ptb01} or tri-polar
structures \citep{pt05}, are not consistent with SAKR properties,
although they may be important in other types of AKR fine structures.

If SAKR bursts are in fact triggered by ion holes, a number of derived
properties of ion holes can be deduced which would be difficult or
impossible to obtain using {\it in situ} measurements. These include
average lifetimes (a few seconds), evolution of propagation speed 
(nearly constant over lifetime of hole), and relative numbers versus
location
(much more common high in the acceleration region
than near the base). 

\begin{acknowledgements} We are grateful to Iver Cairns and Jolene
Pickett for several useful discussions.  This research is supported by
NASA GSFC grant NNG04GB986 and NSF grant ATM 04-07155.
\end{acknowledgements}

\end{article}

\begin{thebibliography}{}

\bibitem[{\it Baumback and Calvert}(1987)]{bc87} 
Baumback, M. and Calvert, W., The Minimum Bandwidths of AKR, 
{\it Geophys. Res. Lett.}, Vol. 14, 119-122, 1987.

\bibitem[{\it Begelman et al.}(2005)]{b05} 
Begelman, M, R. Ergun, and M. Rees, Cyclotron maser emission from blazer jets?
\textit{Astrophys. J.}, 625:51-59, 2005.

\bibitem[{\it Benson and Fainberg}(1991)]{bf91} 
Benson, R. F. and J. Fainberg, Auroral The maximum power of auroral
kilometric radiation
\textit{J. Geophys. Res.}, 96,A8, 13749-13762, 1991.

\bibitem[{\it Benson et al.}(1988)]{b88}
Benson, R., M. Mellott, R. Huff, and D. Gurnett, Ordinary mode auroral
kilometric radiation fine structure observed by DE 1, 
\textit{J. Geophys. Res.}, Vol. 93, xx, 7715-xx, 1988.

\bibitem[{\it Bingham and Cairns}(2000)]{b00}
Bingham, R. and R. Cairns, 
Generation of auroral kilometric radiation by electron horseshoe
distributions,
\textit{Phys. Plasmas}, Vol. 7, 7, 3089-3092, 2000.

\bibitem[{\it Bounds et al.}(1999)]{b99} 
Bounds S. R., et al. Solitary Structures Associated with Ion and Electron 
Beams near 1 Re Altitude, {\it J. Geophys. Res}., vol. 104, 
A12, 28709-28717, 1999.

\bibitem[{\it Calvert}(1982)]{c82} 
Calvert, W., A feedback model for the source of auroral kilometric
radiation,
\textit{J. Geophys. Res.}., 87, 8199, 1982.

\bibitem[{\it Calvert}(1987)]{c87} 
Calvert, W., Hollowness of the observed auroral kilometric radiation
pattern,
\textit{J. Geophys. Res.}., 92, A2, 1267-1270, 1987.

\bibitem[{\it Crumley et al.}(2001)]{c01} 
Crumley, J, C. Cattell, R. Lysak, and J. Dombeck, Studies of ion 
solitary waves using simulations including hydrogen and oxygen beams, 
\textit{J. Geophys. Res.}, 106, A4, 6007-6015, 2001.

\bibitem[{\it DeLory et al.}(1998)]{d98} 
De Lory G. T., et al. FAST Observations of Electron Distributions Within AKR 
Source Regions, \textit{Geophys. Res. Lett.} Vol. 25, 12, 2069-2072, 1998.

\bibitem[{\it Dombeck et al.}(2001)]{d01} 
Dombeck, J., e al., Observed trends in auroral zone ion mode solitary 
wave structure characteristics using data from Polar, 
\textit{J. Geophys. Res.}, 106, A9, 19013-19021, 2001.

\bibitem[{\it Eliasson and Shukla}(2005)]{es05}
Eliasson, B. and Shukla, P., The dynamics of electron and ion holes,
\textit{Nonlin. Proc. Geophys.}, 12, 269 - 289, 2005.

\bibitem[{\it Ergun et al.}(1998)]{e98} 
Ergun R. E., et al., FAST Satellite Wave Observations in the 
AKR Source Region, \textit{Geophys. Res. Lett}., Vol. 25, 12, 2061-2064,
1998.


\bibitem[{\it Ergun et al.}(2000)]{e00} 
Ergun, R. E., et al., Electron-Cyclotron Maser Driven by 
Charge-Particle Acceleration From Magnetic Field-Aligned Electric Fields, 
\textit{Astrophys. J.}, vol. 538, 456-466, 2000.

\bibitem[{\it Farrell}(1995)]{f95} 
Farrell, W., Fine structure of the auroral kilometric radiation: A Fermi
acceleration process?
\textit{Radio Sci.}, Vol. 30, 4, 961-973, 1995.

\bibitem[{\it Farrell et al.}(2004)]{f04} 
Farrell, W. et al., The radio search for extrasolar planets with LOFAR,
\textit{Plan. Sp. Sci.}, Vol. 52, 1469-1478, 2004.

\bibitem[{\it Frank et al.}(2001)]{f01} 
Frank, C., T. Klinger, A. Piel, and H. Schamel, Dynamics of periodic ion
holes in a forced beam-plasma experiment,
\textit{Phys. Plasmas}, 8,10, 4271-4274, 2001.

\bibitem[{\it Grabbe}(1982)]{g82}
Grabbe, C., Theory of Fine Structure of Auroral Kilometric Radiation, 
\textit{Geophys. Res. Lett.}, Vol. 9, Nr. 2, 155-158, 1982.

\bibitem[{\it Green and Gallagher}(1985)]{gg85} 
Green J. L. and Gallagher, D. L., The Detailed Intensity Distribution of the 
AKR Emission Cone, \textit{J. Geophys. Res.,} Vol. 90, A10, 9641-9649, 1985.

\bibitem[{\it Gurnett et al.}(1979)]{g79} 
Gurnett, D. et al., Initial Results form the ISEE-1 and 2 Plasma Wave Investigations, 
\textit{Space Sci. Rev.}, Vol. 23, 103, 1979.

\bibitem[{\it Gurnett and Anderson}(1981)]{ga81} 
Gurnett, D. and R. Anderson, The Kilometric Radio Emission Spectrum:
Relationship to Auroral Acceleration Processes,
\textit{Physics of Auroral Arc Formation}, Geophysical Monograph Series, Vol. 25.
Ed. S.-I. Akasofu and J.R. Kan. Washington DC: American Geophysical
Union, 341-350, 1981.

\bibitem[{\it Gurnett et al.}(1997)]{g97} 
Gurnett, D, R. Huff, and D. Kirchner, The wide-band plasma wave investigation, 
Space Sci. Rev., 79, 195-208, 1997.

\bibitem[{\it Hanasz et al.}(2001)]{h01} 
Hanasz, J., et al., Wideband bursts of auroral kilometric radiation and
their association with UV auroral bulges,
\textit{J. Geophys. Res.}, vol. 106, A3, 3859-3872, 2001.

\bibitem[{\it Kellett et al.}(2002)]{k02} 
Kellett, B., R. Bingham, R. Cairns, and V. Tsikoudi, Can Late-type stars
be explained by a dipoles magnetic trap?
\textit{M.N.R.A.S.}, 239, 102-108, 2002.

\bibitem[{\it Kumamoto et al.}(1998)]{k98} 
Kumamoto, A. and H. Oya, Asymmetry of Occurrence Frequency 
and Intensity of AKR between summer polar region and winter polar region, 
\textit{Geophys. Res. Lett}. Vol. 25, 2369-2373, 1998. 

\bibitem[{\it Louarn, Le Qu\'{e}au, and Roux}(1986)]{l86}
Louarn, P., D. Le Qu\'{e}au, and A. Roux, A new mechanism
of stellar radiobursts: The fully energetic electron maser, 
\textit{Astron. Astrophys.}, vol. 165, 211-217, 1986.

\bibitem[{\it Louarn et al.}(1990)]{l90}
Louarn, P., A Roux, H. de Feraudy, D. Le Qu\'{e}au, M. Andre, and L. Matson,
Trapped electrons as a free energy source for the auroral kilometric
radiation, \textit{J. Geophys. Res.}, Vol. 95, 5983-5995, 1990.


\bibitem[{\it McFadden et al.}(2003)]{m03} 
McFadden J. P., et al., FAST Observations of Ion Solitary 
Waves, \textit{J. Geophys. Res}., Vol. 108, A4, 8018, 2003.

\bibitem[{\it McKeen and Winglee}(1991)]{mw91} 
McKean, M. and R. Winglee, A model for the frequency fine structure of
auroral kilometric radiation,
\textit{J. Geophys. Res.}, 96, A12, 21055-21070,1991.

\bibitem[{\it Menietti et al.}(1996)]{m96} 
Menietti D. et al., Discrete, Stimulated AKR Observed in the Galileo and DE 1 
wideband Data,\textit{ J. Geophys. Res. }Vol. 101, A5, 10673-10680, 1996.

\bibitem[{\it Menietti et al.(2000)}]{mppg00} 
Menietti, D., Persoon, A., Pickett, J., and Gurnett, D., Statistical 
Studies of AKR fine Structure Striations Observed by Polar, 
\textit{J. Geophys. Res.,} Vol. 105, A8, 18857-18866, 2000.

\bibitem[{\it Morozova et al.(2002)}]{mm02} 
Morozova, E., M. Mogilevsky, J. Hanasz, and A. Rusanov,
The Fine Structure of the AKR Electromagnetic Field as Measured by the
Interball-2 Satellite,
\textit{J. Cosmic Res.,} Vol. 40, 4, 404-410, 2002.

\bibitem[{\it Muschietti et al.}(2002)]{m02} 
Muschietti, L. et al., Modeling stretched solitary waves along magnetic field lines, 
\textit{Nonlin. Proc. Geophys}., 9, 101, 2002.

\bibitem[{\it Omidi and Gurnett}(1982)]{og82} 
Omidi N. and Gurnett, D. A., Growth Rate Calculations of Auroral Kilometric 
Radiation Using the Relativistic Resonance Condition, 
\textit{J. Geophys., Res.,} Vol. 87, A4, 2377, 1982.

\bibitem[{\it Pottelette et al.}(2001)]{ptb01} 
Pottelette, R., R. Treumann, and M. Berthomier, Auroral Plasma Turbulence 
and the cause of the AKR fine structure, 
\textit{J. Geophys. Res}. 106, A5, 8465-8476, 2001.

\bibitem[{\it Pottelette et al.}(2003)]{ptbj03} 
Pottelette R., Treumann, R. A., Berthomier, M., and Jaspers, J., 
Electrostatic Shock Properties Inferred from AKR Fine Structure, 
\textit{Non. Proc. Geophys}, Vol. 10, 87-92, 2003.

\bibitem[{\it Pottelette and Treumann}(2005)]{pt05} 
Pottelette R. and Treumann R. A., Electron Holes in the Auroral Upward 
Current Region, \textit{Geophys. Res. Lett}. , Vol. 32, L12104, 2005.

\bibitem[{\it Pritchett}(1984)]{p84} 
Pritchett P. L., Relativistic dispersion, the cyclotron maser instability,
and auroral kilometric radiation,
\textit{J. Geophys. Res}., Vol. 89, 8957-8970, 1984.

\bibitem[{\it Pritchett et al.}(1999)]{p99} 
Pritchett P. L., et al., Free Energy Sources and Frequency 
Bandwidth for the Auroral Kilometric Radiation, 
\textit{J. Geophys. Res}., Vol. 104, A5, 10317-10326, 1999.

\bibitem[{\it Pritchett et al.}(2002)]{p02} 
Pritchett P. L., R. Strangeway, R. Ergun, and C. Carlson., 
Generation and propagation of cyclotron maser emissions in the finite
auroral kilometric radiation source cavity,
\textit{J. Geophys. Res}., Vol. 107, A12, doi:10.1029/2002JA009403, 2002.

\bibitem[{\it Roux et al.}(1993)]{r93} 
Roux A. et al., Auroral Kilometric Radiation Sources: In Situ and remote 
Observations From Viking, \textit{J. Geophys. Res}., Vol. 98, A7, 
11657-11670, 1993.

\bibitem[{\it Speirs et al.}(2005)]{s05} 
Speirs, D., et al., A laboratory experiment to investigate auroral
kilometric radiation emission mechanisms,
\textit{J. Plasma Phys..,} 71, Part 5, 665-674, 2005.

\bibitem[{\it Strangeway et al.(1998)}]{s98} 
Strangeway, R. J. et al., FAST observations of VLF waves in the auroral
zone: Evidence of very low plasma densities,
\textit{Geophys. Res. Lett.} 25, 12, 2065-2068, 1998.

\bibitem[{\it Strangeway et al.(2001)}]{s01} 
Strangeway, R. J. et al., Accelerated Electrons as the Source of AKR,
\textit{ Phys. Chem. Earth (C),} 26, 145-149, 2001.

\bibitem[{\it Temerin et al.(1982)}]{t82} 
Temerin, M. et al. , Observations of double layers and solitary waves in 
auroral plasmas, \textit{Phys. Rev. Lett.,} 48, 1175, 1982.

\bibitem[{\it Willes and Wu}(2004)]{w04} 
Willes, A. and K. Wu, Electron-cyclotron maser emission from white dwarf
pairs and white dwarf planetary systems,  
{\textit M.N.R.A.S.}, Vol. 348, 285-296, 2004.

\bibitem[{\it Wu and Lee}(1979)]{w79} 
Wu, C.S. and Lee, L. C., A Theory of the Terrestrial Kilometric Radiation, 
{\textit Astrophys. J.}, Vol. 230, 621-626, 1979.

\bibitem[{\it Yoon and Weatherwax}(1998)]{y98} 
Yoon, P. and A. Weatherwax, A theory for AKR fine structure, 
\textit{Geophys. Res. Lett}., 25, 24, 4461-4464, 1998.

\bibitem[{\it Zarka}(1998)]{z98} 
Zarka, P., Auroral Radio Emissions at the outer planets: Observations and
theories, 
\textit{J. Geophys. Res.}., 103, E9, 20,159-20194, 1998.

\end{thebibliography}
\end{document}